\documentclass[]{Liebert_Author}
\usepackage{multirow}
\usepackage[table,xcdraw]{xcolor}
\usepackage{colortbl}
\usepackage{graphicx}
\usepackage{float}
\usepackage[caption = false]{subfig}
\usepackage{natbib}
\usepackage[
    starfontserif 
    ]{starfont}
\usepackage{amsmath}
\usepackage{subscript}

\DeclareSymbolFont{starfontsym}{OT1}{sts}{m}{n}
\DeclareMathSymbol{\mathSun}{\mathord}{starfontsym}{115}
\DeclareMathSymbol{\mathTerra}{\mathord}{starfontsym}{76}
\DeclareMathSymbol{\mathvarTerra}{\mathord}{starfontsym}{108}

\title{Dearth of Photosynthetically Active Radiation Suggests No Complex Life on Late M-Star Exoplanets}

\author{Joseph J. Soliz$^{1}$ and William F. Welsh$^{1}$ \\
{$^{1}$Department of Astronomy, San Diego State University}\\
{San Diego, USA}\\
}

\begin{document} 

\maketitle 

\begin{abstract}

The rise of oxygen in the Earth's atmosphere during the “Great Oxidation Event” (GOE) is fairly
well understood to have occurred $\sim$2.3 billion years ago. There is considerably greater uncertainty in the date for the origin of oxygenic photosynthesis, but most studies suggests it occurred significantly earlier, perhaps $\sim$700 million years earlier. 
Assuming this time lag is proportional to the rate of oxygen generation, we can estimate how long it would take for a GOE-like event to occur on a hypothetical Earth-analog planet orbiting 
the star TRAPPIST-1 (an ultra-cool M star with T$_{\mathrm{eff}} \sim$2560 K).
Despite being in the habitable zone, a hypothetical Earth located in TRAPPIST-1e's orbit
would receive only $\sim$0.9\% of the “Photosynthetically Active Radiation” (PAR) that the Earth 
gets from the Sun because most of the star’s energy is emitted at wavelengths longer than the 400-700 nm PAR range.
Assuming oxygen production is proportional to the number of PAR photons, it would take $\sim$63 Gyrs for a GOE, and a staggering $\sim$235 Grys for a Cambrian Explosion.
But the linear assumption is problematic: as light levels increase, photosynthesis saturates then declines -- an effect known as photoinhibition. Photoinhibition varies from species to species and depends on a host of environmental factors. 
There is also high sensitivity to the upper wavelength limit of the PAR for such a red star: Extending just 50 nm increases the number of photons by a factor of $\sim$2.5.
Including these and other factors greatly reduces the timescale to $\sim$1-5 Gyrs for a GOE, and $\sim$4-13 Gyrs for a Cambrian Explosion.
However, non-oxygenic photosynthetic bacteria can thrive in low-light environments and
use of near-IR photons out to $\sim$1100 nm. This provides 22 times as many photons for anoxygenic photosynthesis than oxygenic.
With this huge light advantage, and because they evolved earlier, it is likely that anoxygenic photosynthesizers would dominate the biosphere of a TRAPPIST-1e Earth-analog planet. 
We conclude that on such a hypothetical planet, oxygen would never reach significant levels in the atmosphere, let alone a Cambrian Explosion. Thus complex animal life on such planets is very unlikely.

\end{abstract}


\section{Introduction}

The origin of oxygenic photosynthesis is arguably the most important event in the history of life on Earth, apart from the origin of life itself. The presence of free oxygen in the atmosphere profoundly affected the trajectory of life, and was likely a prerequisite for the evolution of complex multicellular organisms such as plants and animals 
(e.g.\ see \citet{Catling_2005, Ward_2019} and references therein).
Starting from near-negligible amounts in the Hadean and early Archean eons, atmospheric oxygen did not become significant until the ``Great Oxidation Event'' (GOE) roughly 2.3 Ga (e.g.\ see \cite{Luo_2016, ward_timescales_2016, Catling_Kasting_2017, Catling_Zahnle_2020}). While the date of the GOE is fairly well constrained by geological evidence (e.g.\ see \cite{lyons_rise_2014, wang_two-billion-year_2025}),
the date of the origin of oxygenic photosynthesis is not. 
However, several lines of evidence point to an origin well before the GOE 
(\cite{crowe_atmospheric_2013, planavsky_evidence_2014, lyons_rise_2014, 
cardona_PSII_2019, Sánchez-Baracaldo_2020, Fournier_2021}),
implying a delay of perhaps $\sim$700 million years between the origin of oxygenic photosynthesis and the GOE.  An additional $\sim$1.7 billion years would pass before the Cambrian Explosion and the rise of complex multi-cellular life. These are very significant time delays.

Oxygenic photosynthesis depends on photons in the 400-700~nm wavelength range -- the so-called photosynthetically active radiation (PAR) range (e.g.\ see \cite{Blankenship_2014}).
This coincides with peak of the solar radiation spectrum, 
and roughly 22\% of the Sun’s photons are in this range. 
For habitable zone planets orbiting cooler stars, where the spectrum peaks further to the red/infrared, the number of PAR photons will be significantly fewer than what the Earth receives,
and this can have a profound influence on the evolution of life
(e.g.\ see \cite{WOLSTENCROFT2002535, Kiang_2007_I, Kiang_2007_II, Gale_Wandel_2017, Lehmer_2018, Ritchie_Larkum_Ribas_2018, covone_efficiency_2021}). 
While oxygenic photosynthesis can occur under surprisingly dim low-light conditions, 
such as deep
underwater or in caves (\citet{Hawes_Schwarz_2001, Behrendt_2015.14, averina_far-red_2018, Behrendt_2020}), the lower photon flux will, in general, result in a lower oxygen production rate. 
Thus the timescale to produce a global oxygen-rich atmosphere (i.e.\ a GOE) on a planet where such low-light conditions are ubiquitous can be much longer than the $\sim$700 Myr it took on Earth, with far-reaching consequences for the development of complex life on such a planet. 
In this paper, we explore this above scenario, i.e., the timescales and their implication for life on a planet orbiting a low-temperature star. 

Approximately $\sim$80\% of the 100-400 billion stars in our galaxy are the M-type red dwarf stars (\cite{Lada_2006}), and it is estimated that $\sim$16\% of these stars host an Earth-size planet (1–1.5 R$_{\mathTerra}$) in the habitable zone (\cite{Dresssing_Charbonneau_2015}).
To be both specific and extend to the low-temperature limit, we use the TRAPPIST-1 system for
our investigation since the star is an ultra-cool M star star with several Earth-size planets in the habitable zone (\cite{gillon_temperate_2016, gillon_seven_2017}). In particular, we will focus on the planet TRAPPIST-1e.
\citet{Lehmer_2018} examined the TRAPPIST-1 system, finding that for the outermost planets the oxygenic photosynthesis productivity would be limited by the incident light, not by nutrients (as is the case on Earth). 
For these planets the amount of PAR light would likely be insufficient to support an Earth-like biosphere (and thus unable to produce a measurable biosignature). 
Planet e could (barely) receive enough PAR photons for oxygen production to overcome oxygen sinks and build up in the atmosphere, but it could potentially take billions of years. In this paper, we examine this in greater detail.

Despite the ongoing work attempting to characterize the atmospheres of the TRAPPIST-1 planets using \textit{James Webb Space Telescope} (JWST) transit spectroscopy (\citet{greene_thermal_2023,zieba_no_2023,Piaulet-Ghorayeb_2025, Espinoza_2025, Glidden_2025}) our knowledge of such systems remains extremely limited, i.e.,
even if a substantial atmosphere were detected, we would not have information on the crustal composition, tectonics, internal heating,  ocean chemistry, etc. This severely hampers attempts to draw any conclusions about life on these specific worlds. 
We can nonetheless explore a related intriguing question: What would happen if the TRAPPIST-1e planet were replaced with the nascent Earth? How would life on such a hypothetical planet evolve? 
We of course cannot answer this question, but we can look at a few of the key defining events in the history of life on Earth and examine in a general sense how things would differ on a hypothetical Earth-analog planet. Such an exploration may provide some insight into an otherwise intractable problem.

To proceed, we make the ansatz that the only difference between this hypothetical planet and the actual Earth is the spectral energy distribution of the incident radiation from the host star. The crux of our question is thus, ``How would replacing the Sun's spectrum with TRAPPIST-1's spectrum affect the development of life on Earth?''. We proceed by first comparing the TRAPPIST-1 star with the Sun, and estimate the timescales for a GOE. We start with the assumption that oxygen production is directly proportional to the number of PAR photons, then attempt to use the actual oxygen production rates as measured in cyanobacteria and other relevant species. We examine various factors that limit the precision of our timescale estimates, and conclude with discussion of the results.


\section{TRAPPIST-1e}

TRAPPIST-1 is an ultracool, very low-mass star, just above the brown-dwarf mass limit: 
T$_{\mathrm{eff}}$=2566~K, L=0.0005 $L_\odot$, M=0.09 $M_\odot$ (\cite{ gillon_seven_2017, Mann_2019, ducrot_e_trappist-1_2020, Agol_2021}). 
Remarkably, the star hosts seven transiting terrestrial-size planets, three of which reside in the optimistic habitable zone (HZ). 
Located only $\sim$12 pc away, it is a relatively bright system, enabling high signal-to-noise observations. 
These features make TRAPPIST-1 one of the most interesting exoplanet systems currently known.
Of the three planets in the habitable zone, TRAPPIST-1e is particularly interesting, as its size and location (R=0.920  R$_{\mathTerra}$, M=0.692  M$_{\mathTerra}$, a=0.02925 au) makes it quite Earth-like  (\cite{Kopparapu_2013, Lincowski_2018, Agol_2021}).
Its orbit results in an equilibrium temperature of 250 K (-23.5 degrees C) (\cite{ducrot_e_trappist-1_2020}), as it receives 65\% of the Sun-Earth incident radiative flux (\cite{Agol_2021}). 
\citet{Lincowski_2018} provide a UV-optical-IR spectrum of TRAPPIST-1 by combining
scaled empirical observation, a {\textsc{PHOENIX}} stellar spectral model, and a blackbody.
We use this to estimate the the incident photon flux spectrum on TRAPPIST-1e. 
Fig.~1 compares the top-of-atmosphere photon flux spectra for TRAPPIST-1e, the modern-day Earth (\cite{nrel_astm_e490}),
and the Archean Earth at 2.65 Ga from \citet{Claire_2012}.
The shift of the bulk of the photons out of the optical PAR for the Sun and into the 
near-infrared for TRAPPIST-1 is readily apparent.
Only 0.15\% of TRAPPIST-1's photons fall into the PAR range, roughly 150 times less than the Sun's.
Accounting for their orbital distances, the number of PAR photons for TRAPPIST-1e is 
only 0.9\% of what is incident on Earth from the Sun. 
This very low fraction immediately tells us that there may be a problem for oxygenic photosynthesis and the subsequent evolution of life on such a planet.


\section{Estimating the Timescale for a Great Oxidation Event}


\subsection{The Origin of Oxygenic Photosynthesis and the Great Oxidation Event on Earth}

Both the timing and the mechanism(s) leading to the origin of oxygenic photosynthesis are topics of are of intense interest. 
The latter is likely to have occurred after the origin of non-oxygenic photosynthesis, as it depends on combining two independent photosystems (PS I and PS II), while anoxic photosynthesis requires only one photosystem (see e.g.\ \cite{Blankenship_2014}). 
Dates for the origin for oxygenic photosynthesis range from just prior to the GOE (\cite{ward_timescales_2016, Shih_2017, Soo_2017, kasting_atmospheric_2025}) 
to $\geq$ 2.7 Ga (based on fossilized rock layers of  stromatolites and from chlorophyll-a biomarkers hosted in shale; e.g.\ see \cite{buick_2008,Sim_2012,KURZWEIL201317,WILMETH2019}) 
to $\geq$ 3.0 Ga (based on geological traces of O$_2$ and on bacterial lineages of photosystems; (e.g.\ see \cite{crowe_atmospheric_2013, planavsky_evidence_2014, ROBBINS2023, patry_dating_2025, wang_mesoarchean_2018, cardona_PSI_2018, cardona_PSII_2019,Fournier_2021, boden_timing_2021}).
Sporadic ``whiffs'' of oxygen may also provide evidence for an early rise of oxygenic photosynthesis (\cite{Anbar_whiff_oxygen_2007}) though this is disputed (e.g.\ \cite{Slotznick_reexamne_whiffs_2022}).
We adopt an origin of of oxygenic photosynthesis at 3 Ga, cognizant that there is considerable uncertainty in this value.

The date of the Great Oxidation Event is much better determined, based on the appearance of red banded iron formations, 
the disappearance of detrital pyrite grains in sediments, 
and distinctive mass-independent sulfur isotope fractionations (\cite{Luo_2016, lyons_rise_2014}). 
These point to a time between 2.4 to 2.22 Ga (\cite{Luo_2016, Catling_Zahnle_2020, poulton_200-million-year_2021, Lyons_2021}), and we adopt 2.3 Ga for the GOE.
The time lag between the origin of oxygenic photosynthesis and the GOE, $\sim$ 700 million years, is thought to be due to presence of a number of oxygen sinks in the Archean Eon, i.e., O$_2$ production is balanced by consumption and reactions with reducing compounds (e.g.\ the decay of organic matter). But over geologic timescales, the burial of organic carbon and other reducing agents (particularly hydrogen, sulfur, and iron) resulted in the loss of this sink channel, and tipped the balance allowing oxygen to accumulate in the atmosphere (see \cite{Catling_Kasting_2017} for a thorough discussion).  Other hypotheses for the time lag include the lack of sufficient nutrients, the low efficiency of oxygen production by early protocyanobacteria (e.g., see \citep{Grettenberger_Sumner_2024}),
and even sporadic collisions with large impactors from space resulting in transient but significant oxygen sinks (\cite{marchi_delayed_2021}).

After the rapid rise in atmospheric oxygen levels during the GOE, including a large ``overshoot'' and plunge back to lower levels 
(the Lomagundi excursion; see \cite{lyons_rise_2014}), oxygen levels throughout the Proterozoic Eon fluctuated sporadically around a few thousandths to a few percent of the present day value (\citep{lyons_rise_2014, fischer_evolution_2016, Lyons_2021, Luo_2016, Krause_2022}).
Roughly 1.7 billion years would pass before oxygen levels rose again at the end of 
the Proterozoic, and eventually reaching present day levels. This second rise, known as the Neoproterozoic Oxidation Event (NOE), occurred roughly 860-540 Mya, and preceded the Cambrian Explosion $\sim$540 Ma (\cite{Sperling_2013, fox_what_2016}) at the start of the Phanerozoic Eon. The Cambrian Explosion - sometimes referred to as the ``Big Bang of biology'' - marks the onset of a rapid increase in the diversity and complexity of life on Earth. The question of whether the NOE directly caused the Cambrian Explosion has long been debated, but it seems very likely that the presence of atmospheric oxygen was a necessary factor (see e.g.\ \cite{Catling_2005, sperling_ecological_2015, Ward_2019}, also see \cite{fox_what_2016} for a very readable review).

\subsection{Simple Linear Approximation}
To estimate how long it would take for a GOE-like event to happen on our hypothetical TRAPPIST-1 Earth-analog planet, we make the simple assumption that the amount of oxygen produced via oxygenic photosynthesis is directly proportional to the number of surface PAR photons.
Since this hypothetical planet is by construction identical to the Earth, the sinks of oxygen are taken to be the same. Then the timescale between the origin of oxygenic photosynthesis and the GOE is given by
$  \tau_{T} = \tau_{E} \cdot ( {\frac{N_{T}}{N_{E}}} )^{-1} $
where
  $\tau_{E}$ is the timescale for the GOE on Earth ($\sim$700 Myr);
  $N_{E}$ is the PAR photon flux on Earth (number of $\mu$mol photons m$^{-2}$ s$^{-1}$ at the surface); 
  and 
  $N_{T}$ is the PAR photon flux on our TRAPPIST-1e Earth-analog planet.
For the photon flux number, we use the Archean solar spectrum not the present-day spectrum,
as shown in Fig.~1. For a late M-type star like TRAPPIST-1, the small PAR photon flux relative to the Earth will result in a significantly delayed onset for the GOE. For TRAPPIST-1, the timescale for a GOE is roughly $\sim$63 Gyrs, much older than the universe. The time to a Cambrian Explosion event would take a staggering $\sim$235 Gyrs.
Although this is a crude estimate, it nevertheless suggests that the path life takes on a planet orbiting a late M star will likely be very different than what occurred on Earth.


\subsection{Improving the Timescale Estimate}
A variety of factors affect the GOE timescale on our Earth-analog planet, some precise and easy to calculate, others quite confounding. We explore these below.

\subsubsection{Far-Red Light and the Extended PAR}
Oxygenic photosynthesis can use five types of chlorophyll (\citet{Blankenship_2014}).
Plants usually use chlorophyll a and b, and algae typically use chlorophyll c, but some cyanobacteria can also use chlorophyll d and f. These cyanobacteria can use light beyond the “red limit” of the standard PAR, extending the range out to 750 nm 
(e.g.\ see \cite{Kiang_2007_I, Kiang_2007_II,  Kiang_2022, Chen_2010, mielke_2013, Ritchie_Larkum_Ribas_2018, Schwieterman_2018, antonaru_global_2020}). 
On Earth, extending the PAR by 50 nm is not of great consequence: it adds $\sim$17\% more photons. But for a star like TRAPPIST-1, whose spectrum peaks in the NIR, this small extension 
{\it increase the number of photons available for oxygenic photosynthesis by $\sim$250\%} 
(see Fig.\ 1). 
If only the incident fluxes are considered, this would 
reduce the timescale between the origin of oxygenic photosynthesis and the GOE 
by a factor of $\sim$2.

Chlorophyll d and f are not used by many species on Earth, but with such an abundance of extra photons available on a planet orbiting TRAPPIST-1, one would expect evolutionary pressures to push towards its use. \textit{Acaryochloris marina} is a low-light adapted cyanobacteria that uses chlorophyll-d as its primary pigment to absorb photons, extending the PAR out to
$\sim$720-750 nm  (\cite{Kiang_2007_I, Kiang_2022, mielke_2013,ritchie_fitting_2008}), 
and thus is potentially the cyanobacteria best equipped to cope with a TRAPPIST-1e-like environment. The cyanobacteria \textit{Halomicronema hongdechloris} 
will produce chlorophyll-f when PAR photons are scarce, extending its light-harvesting into the far-red 
($\sim$ 740-760 nm) (\cite{Chen_2010, antonaru_global_2020}). 
\citet{Behrendt_2015.14, Behrendt_2020} found that cyanobacteria with chlorophyll-f can thrive deep in caves where the low-light environment is enriched in far-red and NIR light caused by reflectance off the surface materials of cave walls. 
\citet{Battistuzzi_2023} experimented with the cyanobacteria \textit{Chlorogloeopsis fritschii PCC6912} and \textit{Synechocystis sp. PCC6803} under conditions designed to simulate the spectra of M stars. Both species thrived, with \textit{Synechocystis} being efficient with the photons in the PAR, while \textit{C. fritschii} acclimated to the far-red light conditions and harvested both visible and far-red light using chlorophyll-d and f.
These results make it clear that the standard PAR should be extended beyond the traditional red limit and that an extended PAR, spanning 400-750 nm, should be used for this investigation.


\subsubsection{Synchronous Rotation}

With an orbital period of 6.1~d (\cite{Agol_2021}), TRAPPIST-1e is likely to experience tidal locking (i.e., synchronous rotation), creating perpetually bright and dark hemispheres (see the seminal paper by \cite{kasting_habitable_1993} for a discussion). 
However, continuous illumination is not likely to be an impediment to the ability for life to thrive (see e.g.\ \cite{Hu_Yang_2014, Gale_Wandel_2017, Lobo_2023}). 
For a synchronously rotating planet, the day-side hemisphere will receive twice as many photons on average than a hemisphere experiencing day-night cycles. However, any oxygenic photosynthesis will only occur on the day-side hemisphere, reducing the global surface biomass by a factor of two. 
To first order, these two effects cancel. We are ignoring the apparent correlation between rotation period and net atmospheric oxygen on Earth (slower rotation leads to more atmospheric oxygen; see \cite{klatt_rotation_2021}), because the validity of extrapolating slowing rotation to no rotation (i.e.\ synchronous) is not obvious.


\subsubsection{Photoinhibition and Photosynthesis-Irradiance (PI) Curves}

Oxygen production is highly non-linear due to a well-known effect called photoinhibition. At ``high light levels'', oxygen production saturates then declines (e.g.\ see \cite{kok_inhibition_1956,  Falkowski_Raven_2007}). 
Accounting for photoinhibition is thus important, but it is challenging task because the rate of oxygen generation for a given incident light level depends on many factors such as temperature, pH, salinity, and the availability of nutrients. 
More importantly, different species of cyanobacteria can have very different oxygen production rates, and even within a subspecies the process of photoacclimation can alter the response when exposed to persistent bright or dim light. 
For example, some species will shift from using chlorophyll a to d or f when in chronic low-light conditions (e.g.\ \cite{Battistuzzi_2023}). 
Thus the quantitative meaning of ``high light levels'' is very subspecies and environment specific.

A figure showing the rate of oxygen production versus incident PAR photon flux is known as a “photosynthesis-irradiance (P-I) curve”. Examples for various species of cyanobacteria are shown in Fig.~2. 
[The Supplemental Information provides the parameters used to created these P-I curves.].
Because of the above-mentioned variations in oxygen production rate, these curves have been normalized to their peak, and unlike standard P-I curves we plot the logarithm of irradiance to better show the behavior over the large variation in incident photon flux relevant to late M-stars. For species accustomed to bright light, oxygen production often flattens and declines when light levels exceed the daily average incident photon flux level on Earth’s surface: $\sim$ 800 $\mu$mol PAR photons $m^{-1} s^{-1}$ (\cite{Nobel_2005}).  
However, the Archean Sun 2.65 Ga was $\sim$20\% fainter than the present-day sun (\cite{Claire_2012}),  providing an average PAR photon flux of 652 $\mu$mol photons $m^2 s^{-1}$, and 764 $\mu$mol photons $m^2 s^{-1}$ for the extended PAR. These value is shown as the vertical blue lines in Fig.\ 2. TRAPPIST-1e-like irradiances are 
 7.3 $\mu$mol m$^{-2}$ s$^{-1}$ in the PAR, and 
18.3 $\mu$mol m$^{-2}$ s$^{-1}$ in the extended PAR, and are shown as the red vertical lines.
The Archean photon flux is still much brighter than many low-light environments on today's Earth  (e.g.\ marine/aquatic habitats, sub-ice, caves, canopy shade, etc.).
Cyanobacteria accustomed to low-light environments will saturate at much lower irradiance, and at the  actual Archean photon flux their oxygen production rate will be near-zero; such a ``bright'' environment inhibits growth. 
Thus directly using P-I curves to compare the oxygen production under Archean versus TRAPPIST-1e-like irradiance levels to estimate a GOE timescale is not valid. 
Thus replacing $( {\frac{N_{T}}{N_{E}}} )^{-1} $ with $(\frac{P(I_{T})} {P(I_{E})})^{-1}$ 
where $P(I)$ is the normalized photosynthesis curve at irradiation level (I), will not yield a sensible result for a timescale estimate. 
So to include the photoinhibition effect, we make the assumption that the peak of the P-I curve corresponds to the irradiance level in the environments the cyanobacteria are best adapted to. 
The oxygen production at the peak is then compared to the oxygen production at the irradiance level TRAPPIST-1e receives. This ratio is then used to scale the 700 Myr timescale that occurred on Earth:
$\tau_{T} = \tau_{E} \cdot (\frac{P(I_{T})} {P(I_{max})})^{-1}$ .
Since $P(I)$ is normalized (i.e.\ $P(I_{max}) \equiv 1$), the timescale is simply inversely proportional to the relative height of the P-I curve at the
TRAPPIST-1e Earth-analog irradiance level.

The large differences in the P-I curves for different species results in a large range for the timescale for oxygenation on our hypothetical planet. But as expected, since the TRAPPIST-1e environment is a low light environment, the shortest timescales result from those species that are adapted to low-light levels. 
Of these species we were able to find published P-I data for, the two most relevant are \textit{Prochlorothrix hollandica PCC 9006} and \textit{Acaryochloris marina} (\cite{ritchie_fitting_2008}). 
Both are adapted to low-light conditions, with peak occurring at 81 $\mu$mol m$^{-2}$ s$^{-1}$  and 194 $\mu$mol m$^{-2}$ s$^{-1}$, respectively.
In particular, \textit{Acaryochloris marina} lives in low-light level environments and uses chlorophyll-d (\cite{Kiang_2022, mielke_2013, Ritchie_Larkum_Ribas_2018}), and thus provides our ``best estimate'' value for the timescale for a GOE.

Another important species is \textit{Gloeobacter violaceus} (\cite{KOYAMA2008369}),
because of all the extant species of cyanobacteria,  those of the \textit{Gloeobacter} genus have the most ancient lineage (\cite{larsson_genome_2011, sanchez-baracaldo_origin_2015, Schirrmeister_2015, Fournier_2021}) and are thus most likely to be similar to the cyanobacteria in the Archean. 
The \textit{Gloeobacter} genus have a more primitive light-harvesting mechanism, using phycobilisomes instead of  using thylakoids. 
In particular, the P-I curve for \textit{G. kilaueensis} would be valuable, at it
was discovered in a lava cave in low-light conditions (\cite{saw_cultivation_2013}). 
But in general, \textit{Gloeobacter} are slower-growing than most other cyanobacteria, resulting in longer timescales for a GOE. 
For comparison, the \textit{Synechococcus} genus of cyanobacteria is an abundant, fast-growing, picoplankton and an important primary producer, and more accustomed to high light levels. 
\textit{Synechococcus R-2 PCC 7942} has a peak oxygen production at an incident photon flux of
at 600 $\mu$mol m$^{-2}$ s$^{-1}$ (\cite{ritchie_fitting_2008}).

Our linear estimate for the timescale between the origin of oxygenic photosynthesis and a global GOE on a hypothetical planet is based only on the ratio of the incident PAR flux. The estimate using photoinhibition is based only on the ratio of the oxygen production rates, independent of the incident flux.
In other words, the linear estimate is sensitive to the x-axis of the P-I curve, while the photoinhibition estimate is sensitive only to the y-axis of Fig.\ 2. Since the number of photons in the PAR is of course crucial, the latter estimate is very likely to be an underestimate, possibly by as much as factor of $\sim 40$ since $\frac{N_{T}}{N_{E}}= 0.024$ in the extended PAR. 
Thus the photoinhibition estimate in our discussion should be viewed as an approximate lower-limit on the timescale.


\subsubsection{Revised Timescale Estimates}

Using the extended PAR and considerations of photoinhibition, the timescales between the origin of oxygenic photosynthesis (OP) and the GOE are greatly reduced compared to the linear assumption. 
Examining cases for which we have \mbox{P-I} curve parameters, the shortest timescale comes from the low-light species \textit{Prochlorothrix h.}, which yields a timescale of $\sim$1.4 Gyr.  
For comparison, the bright-light species \textit{Synechococcus R-2 PCC 7942}, yields 8.7 Gyr. 
The timescale for \textit{Acaryochloris marina}, our ``best estimate'' case, yields 3.0 Gyr. These values are shown in Table~1. 
Also shown are the estimated timescales to reach a Cambrian Explosion (CE). 
For these, we used the Proterozoic solar spectra (\cite{Claire_2012}) for the incident photon flux for Earth when estimating the time to the CE.
These timescale range from 5.0 Gyr to 30.5 Gyr, with 10.5 Gyr for our best estimate (\textit{Acaryochloris marina}). Since the date of the origin of oxygenic photosynthesis contains considerable uncertainty,  we also include minimum and maximum values for the timescales $\tau_E$ based on the range of values found in the literature, as discussed earlier. 
The maximum range spans 3.5 to 2.22 Ga for the OP to GOE, while the minimum range spans 2.7 to 2.4 Ga. 
In our opinion, the minimum range likely significantly underestimates the delay, but we include it for completeness.
There is considerable uncertainty in the estimates in Table~1 (including a potentially large bias towards underestimating the timescales, as noted above). The usefulness of the table comes from recognizing that the timescales are typically significantly longer than what occurred on Earth.


\section{Discussion}

Essentially we are asking the question, ``What would happen if we replaced TRAPPIST-1e with the Archean Earth?''
Because TRAPPIST-1 is a late M star with a very red spectrum, this planet would receive 
$\sim$88 times less photosynthetically active radiation than the Earth received from the Sun.
This would greatly slow the oxygen production rate, and assuming the same timescale for oxygen to accumulate in the atmosphere as it did on Earth, this leads to the simple estimate that it would take $\sim$63 Gyrs 
for oxygen to build up to even a few hundredths of percent in the atmosphere.
A more realistic estimate uses a PAR that extends to the far red by 50 nm, and this has a very significant effect, because it increases the number of photons by a factor of 2.5 
for TRAPPIST-1e.
Including the effects of photoinhibition is challenging for a number of reasons, but focusing on low-light accustomed species can again result in a dramatic reduction in the timescale. 
Our best estimate is $\sim$3 Gyrs for a Great Oxidation Event. This is less than the $\sim$7.6 Gyr age of the TRAPPIST-1 system (\cite{Burgasser_2017}), so it is possible that such a hypothetical planet could have experienced a GOE. However, it would take over 10 Gyrs for a Cambrian Explosion-like event to occur, suggesting that complex life (meaning multicellular life at the cm scale or larger)
would be unlikely (e.g.\ see \cite{Dismukes_2001, Catling_2005}).
But there is another major factor to consider, which we discuss below.


\subsection{Anoxygenic Photosynthesis}

Over the $\sim$3.8 billion years life has existed on Earth, only once has oxygenic photosynthesis evolved -- in the cyanobacteria  (e.g.\ see \cite{Blankenship_2014, ward_timescales_2016}). Given numerous examples of convergent evolution we see in nature, 
this uniqueness suggests that the mechanism is particularly challenging to arise. 
This assertion is supported by the intricacy at the heart of oxygenic photosynthesis,
the ``oxygen evolving complex'' (also called the water oxidizing complex), which
essentially requires the use of an elaborate molecular photoelectric ``capacitor'' to 
enable the oxidation of two molecules of water into O$_2$.
To quote \citet{Dismukes_2001},  
{\it{``... the oxidation of water involves a complex, four-electron / four-proton coupled 
oxidation reaction that is thermodynamically the most challenging multielectron reaction in biology.''}}

By contrast, non-oxygenic (anoxygenic) photosynthesis is present in a variety of bacteria:
the purple sulfur and the purple non-sulfur bacteria; the green sulfur bacteria; 
the filamentous anoxygenic phototrophs (formerly called green non-sulfur bacteria); 
the heliobacter; and the choroacidobacteria (\cite{Blankenship_2014}).
Instead of using two photosystems in tandem as required for oxygenic photosynthesis,
anoxygenic photosynthesis employs one of two possible ``reaction centers'',
and thus is far simpler and likely to have evolved before oxygenic photosynthesis 
(e.g.\ see \cite{Blankenship_2014,Dismukes_2001,fischer_evolution_2016}).
Anoygenic photosynthesis uses bacteriochlorophylls, which are similar to, but distinct from, chlorophylls. Since anoxygenic photosynthesis uses H$_2$S 
(or thiosulfate, ferrous iron, elemental sulfur, or H$_2$) instead of water, bacteriochlorophylls are, in general, able to harvest lower-energy photons to enable photosynthesis.
In particular, some purple sulfur and purple non-sulfur bacteria employ
bacteriochlorphyll-b which peaks near $\sim$1020-1040 nm
(and extends out to $\sim$1100 nm) to run their photosystem machinery
(see e.g.\ \cite{drews_bezeichnung_1966, drews_rhodopseudomonas_1966, 
Madigan_Jung_2009, Ritchie_Larkum_Ribas_2018, larkum_living_2018}).
The purple bacteria are a diverse group, with nearly 50 genera,
some of which are extremophilic (grow best at unusual temperature, pH, or salinity);
some purple non-sulfur bacterial can grow without photosynthesis (via fermentation or chemotrophy), and most can fix nitrogen (\cite{Madigan_Jung_2009}).
Their ability to use NIR light makes the purple sulfur bacteria well-suited for planet orbiting a late M star. 
On the other hand, the green anoxygenic photosynthetic bacteria are particularly well-suited for low-light environments because they possess unique light-harvesting antenna complexes known as chlorosomes (\cite{orf_chlorosome_2013}). 
Although their bacteriochlorophylls c, d, or e do not push to longer wavelengths much beyond the extended PAR, their enormous chlorosome antennae allow photosynthesis at very low intensities. 
Specifically, the low-light adapted green sulfur bacteria ({\it{Chlorobaceae}}), known as the brown-colored GSB, can survive in environments with extremely low intensities, such as the Black Sea chemocline (boundary between the oxygenated and the deep anoxic water, located $\sim$100m below the surface), with a light intensity of only  $\sim$ 0.00075--0.0022 $\mu$mol photons m$^{-1}$ s$^{-1}$, roughly 5 orders of magnitude below the Black Sea surface intensity (e.g.\ \cite{Manske_green_bacteria_2005})
Cultured specimens of green sulfur bacterium BS-1 showed detectable photosynthetic activity at 0.015 $\mu$mol photons m$^{-1}$ s$^{-1}$ (\citep{Manske_green_bacteria_2005}).
Even more remarkably, there is evidence for green sulfur bacteria using the red optical tail of the thermal light emitted from hydrothermal vents (\cite{Beatty_hydrothermal_vent_2005}), where
immediately adjacent to the vent the photon flux is similar to that near the Black Sea chemocline.

On a planet orbiting a late M star like TRAPPIST-1 where the spectral energy distribution peaks in the near infrared, purple bacteria employing anoxygenic photosynthesis would have a {\it huge} advantage over their oxygenic counterparts: extending the PAR out to 1100 nm results in
{\it 22 times as many photons available for anoxygenic photosynthesis} than oxygenic 
(55 times as many in the standard PAR; see Fig.~1). Other types of anoxygenic photosynthetic bacteria can use the far-red light that is not absorbed by the purple bacteria, allowing the coexistence of several anoxygenic phototrophs in the same environment. 
Because anoxygenic evolved before oxygenic photosynthesis, there would have been direct competition for light and nutrients (\cite{ozaki_anoxygenic_2019}).
Given the tremendous upper hand in available photons, we hypothesize that anoxygenic photosynthetic bacteria would dominate,
with perhaps a small niche of low-light protocyanobacteria struggling for resources in an anoxic-dominant biome.
A similar conclusion, that anoxygenic photosynthesis could dominate on M-star planets if there were sufficient non-water electron donors, was reached by \citet{Kiang_2007_II} 
in their detailed investigation of photosynthesis on planets orbiting non-solar like stars.
We further speculate that, on planets orbiting stars like TRAPPIST-1, a GOE would never occur, let alone a Cambrian Explosion. Thus complex animal life would not exist.

While TRAPPIST-1 was chosen for this study because it is an extreme case of an ultracool M dwarf star, the results are generalizable, and any habitable zone planets orbiting late M-dwarf stars, such as LP 890-9 (\cite{delrez_l_two_2022}) or SPECULOOS-3 (\cite{gillon_detection_2024})
may share a similar fate. Since late-M stars are the most common type of star in the galaxy, these results may have important implications on the search and expectations for life beyond Earth. 
 
The possibility of photosynthesis on the planet Proxima-b is discussed in the excellent work by \cite{Ritchie_Larkum_Ribas_2018}. Proxima Centauri is a mid-M star (M5.5 V) with an effective temperature of 3050~K, somewhat hotter than \mbox{TRAPPIST-1} (2566~K). 
This seemingly modest increase in temperature is quite significant however, as M star spectra contain molecular bands in addition to atomic electron lines, resulting in a notably more red spectrum for TRAPPIST-1. 
In particular, the planet Proxima-b receives 3\% as many PAR photons as the Sun, compared to 0.9\% for TRAPPIST-1e. 
On Proxima-b, \cite{Ritchie_Larkum_Ribas_2018} estimate there is enough PAR light for oxygenic photosynthesis in a shallow aquatic environment, 
but conditions would not be favorable for anoxygenic photosynthesizers. 
Nevertheless, \cite{Ritchie_Larkum_Ribas_2018} conclude that a substantial aerobic or anaerobic ecology could be possible on Proxima-b.
We believe that for later (cooler) M-type stars, the balance is tipped in favor of an anaerobic planet, 
not just based on the stellar spectrum, but on the enormously long timescales needed to transition from an anaerobic to aerobic world.


\subsection{Caveats and Concluding Thoughts}

Given the results of this study, it worth explicitly listing the fundamental assumptions made:
(1) By employing terrestrial timescales, we are in effect assuming an exact Earth-analog in terms of physical properties: mass, radius, atmospheric composition and transmission, crustal composition / mineralogy, tectonics, oceans, tides, salinity, pH, etc.
The great advantage of such a broad assumption is that nearly all unknown factors scale out of the problem. We are not asking, ``What are these conditions on TRAPPIST-1e?''; that is currently unknowable. But we do know the timescales for the Earth. \
(2) We assume that life will emerge on this planet on a similar timescale that it did on Earth, and at its early stages will follow a similar path that life did on Earth, i.e., simple prokaryotic life that eventually evolves a mechanism for photosynthesis.
Additional assumptions include:
(3) Lyman $\alpha$ radiation has no effect on the timescales; \
(4) Flares from stellar activity do not significantly disrupt the atmosphere; \
(5) The timescales are not dependent on the slowing of the Earth's rotation over the eons, nor does a tidally-locked rotation have much effect; \
(6) Photoinhibition and P-I curves for extant cyanobacteria are similar to (proto)cyanobacteria prior to the GOE and NOE; \
(7) We assume that aerobic respiration is a requirement for complex life.
While we are not assuming that the path that life takes on this hypothetical world would be identical to what occurred on Earth (even if we rewound and played Earth's history over we do not believe it would be the same), 
but we do make the assumption that the timescales are roughly similar.
Importantly, factors of tens of percent in the estimated timescales do not alter the conclusions.

Our timescales are very sensitive to the P-I curve employed. We used P-I
curves for a variety of cyanobacteria, focusing on the low-light species,
but the most appropriate genus to use is \textit{Gloeobacter}, as this is the most ancient lineage of cyanobacteria \cite{larsson_genome_2011, Schirrmeister_2015, Fournier_2021}, 
and the likely the most similar extant species to the earliest oxygenic synthesizers (the proto-Cyanobacteria - see \cite{Grettenberger_Sumner_2024} and references therein).
We were unable to find a published low-light P-I curve for \textit{Gloeobacter}, 
but given its it slow grow compared to other types of cyanobacteria
(\cite{Raven_SanchezBaracaldo_2021}), 
and that the proto-Cyanobacteria were very likely less efficient at photosynthesis 
\cite{Grettenberger_Sumner_2024}, 
the expectation is that this would lead to lengthening the timescale estimates.

Red and NIR light is highly absorbed by water, thus limiting anoxygenic phototrophs to the surface layers (e.g.\ see \cite{larkum_living_2018}).
To zeroth order, our scaling of the timescale for the GOE on Earth includes this,
but the loss of any PAR photons would have the effect of lengthening the timescale to a GOE.
The loss of far red and NIR light would preferentially reduce the advantage of the anoxygenic phototrophs, 
but with a factor of 22 times more photons available for photosynthesis, anoxygenic photosynthesis is still very likely to dominate.
Nonetheless, the lack of any blue/green light from the host star, 
combined with the loss of red-NIR light due to water absorption, 
leaves scant little remaining to be harvested for photosynthesis.
At such extremely low light levels, growth is expected to be extremely slow 
(e.g.\ doubling time of years to decades for bacteria deep in the Black Sea (e.g.\ \cite{Manske_green_bacteria_2005})).
Thus a planet dominated by slowly-growing, photon-starved anoxic microbial mats, confined to shallow water or damp terrestrial environments, may be the most likely expectation for our late M-star Earth-analog planet.

Finally, if future work shows we are incorrect, i.e., abundant oxygen is found in a late M-dwarf exoplanet’s atmosphere, this would be extremely exciting. It would suggest that life has found a way to carry out oxygenic photosynthesis by combining several NIR photons - an astonishing feat.


\section{Acknowledgments}

We thank Nancy Kiang for extremely valuable feedback on an early poster draft of this work, 
and Pramod Gupta and the Virtual Planetary Laboratory for assistance obtaining the Young Sun Model spectra. We thank the late John Hood Jr., for supporting exoplanet research at SDSU.


\bibliographystyle{aasjournal}
\bibliography{References}


\clearpage
\newpage
\section{Tables and Figures}

\begin{table}[h]
\caption{Estimated Timescales for Oxygen Events}
\label{tab:table1}
\resizebox{\columnwidth}{!}{%
\begin{tabular}{|
>{\columncolor[HTML]{FFFFFF}}c 
>{\columncolor[HTML]{FFFFFF}}l 
>{\columncolor[HTML]{FFFFFF}}l 
>{\columncolor[HTML]{FFFFFF}}c 
>{\columncolor[HTML]{FFFFFF}}c 
>{\columncolor[HTML]{FFFFFF}}c |}
\hline
\multicolumn{2}{|c|}{Cyanobacteria species} & \multicolumn{1}{c|}{Timescales} & \multicolumn{1}{c|}{Best Estimate} & \multicolumn{1}{c|}{Minimum} & Maximum \\ 
\hline
\hline

\multicolumn{2}{|c|}{} & \multicolumn{1}{l|}{OP to GOE} & \multicolumn{1}{c|}
 {8.7} & \multicolumn{1}{c|}{3.7} & 15.8\\ \cline{3-6} 
 
\multicolumn{2}{|c|}{\multirow{-2}{*}{\textit{Synechococcus R-2 PCC 7942}}} 
& \multicolumn{1}{l|}{OP to CE} & \multicolumn{1}{c|}
 {30.5} & \multicolumn{1}{c|} {26.7} & \multicolumn{1}{c|}{36.6} \\ \cline{3-6}
\hline

\multicolumn{2}{|c|}{} & \multicolumn{1}{l|}{OP to GOE} & \multicolumn{1}{c|}
 {1.4} & \multicolumn{1}{c|}{0.6} & \multicolumn{1}{c|}{2.6} \\ \cline{3-6} 

\multicolumn{2}{|c|}{\multirow{-2}{*}{\textit{Prochlorothrix hollandica PCC 9006}}} 
& \multicolumn{1}{l|}{OP to CE} & \multicolumn{1}{c|}
 {5.0} & \multicolumn{1}{c|}{4.4} & \multicolumn{1}{c|}{6.0} \\ \cline{3-6}
\hline

\multicolumn{2}{|c|}{} & \multicolumn{1}{l|}{OP to GOE} & \multicolumn{1}{c|}
 {\bf{ 3.0} } & \multicolumn{1}{c|}{1.3} & \multicolumn{1}{c|}{5.5} \\ \cline{3-6} 

\multicolumn{2}{|c|}{\multirow{-2}{*}{{\bf \textit{Acaryochloris marina}}}} 
& \multicolumn{1}{l|}{OP to CE} & \multicolumn{1}{c|}
 {\bf{ 10.5 } } & \multicolumn{1}{c|}{9.2} & \multicolumn{1}{c|}{12.7} \\ \cline{3-6}
\hline
\end{tabular}
}
\end{table}

\clearpage

\newpage
\pagebreak
\newpage

\begin{figure}[hthp]
    \includegraphics[width=1.0\textwidth]{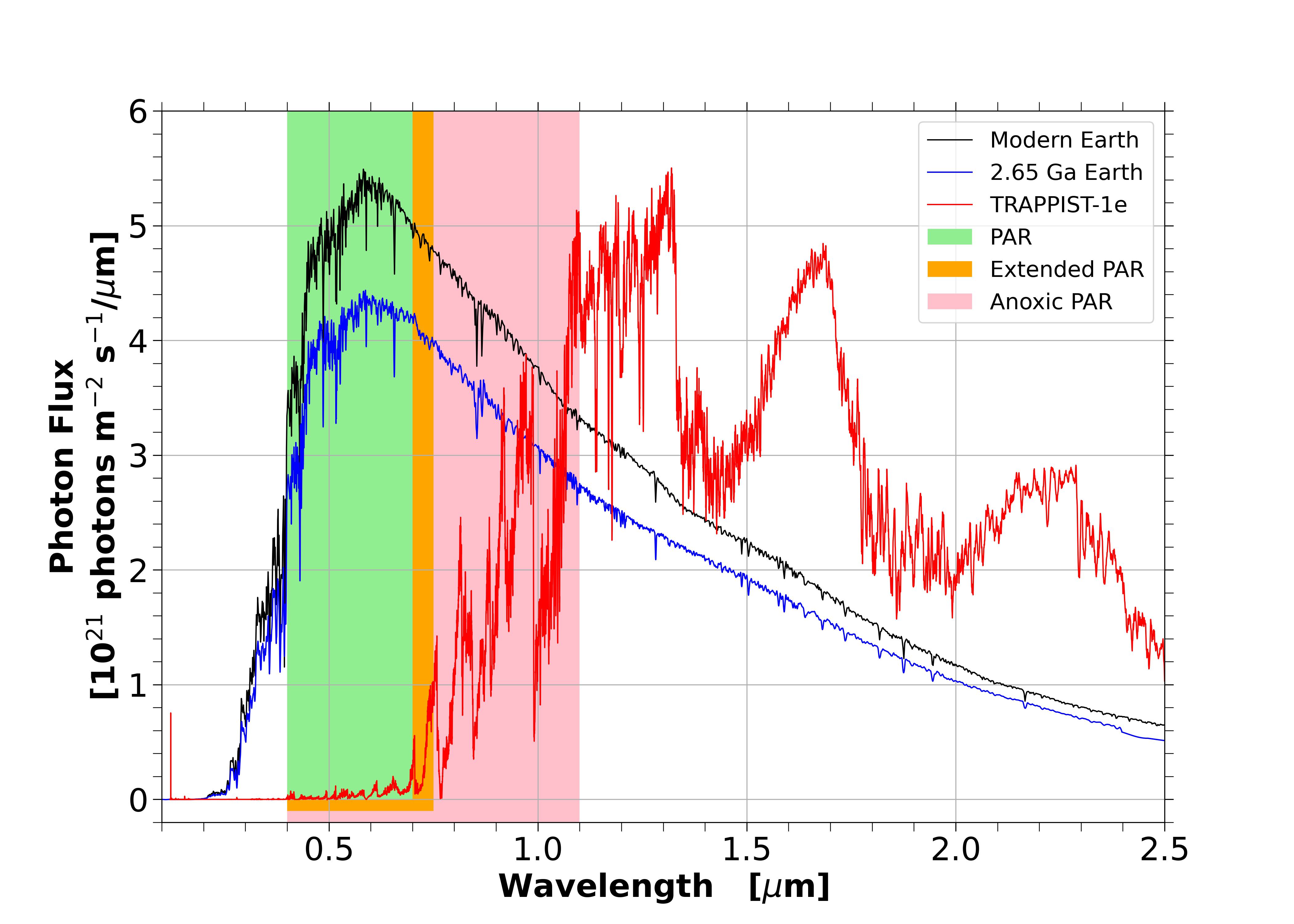}
    \caption{The incident photon flux density for the modern-day Earth (black), Archean Earth at 2.65 Ga (blue), and TRAPPIST-1e (red). The spectral resolution has been reduced for clarity.
    The shaded regions represents three relevant bandpasses for photosynthesis: standard PAR (0.40-0.70 $\mu$m), extended PAR (0.40-0.75 $\mu$m), and anoxic PAR (0.40-1.1 $\mu$m). 
    \label{fig:spectra}}
\end{figure}
\clearpage

\newpage
\pagebreak

\begin{figure}[htbp]
\centering
\subfloat[]{\includegraphics[width = 0.8\textwidth]{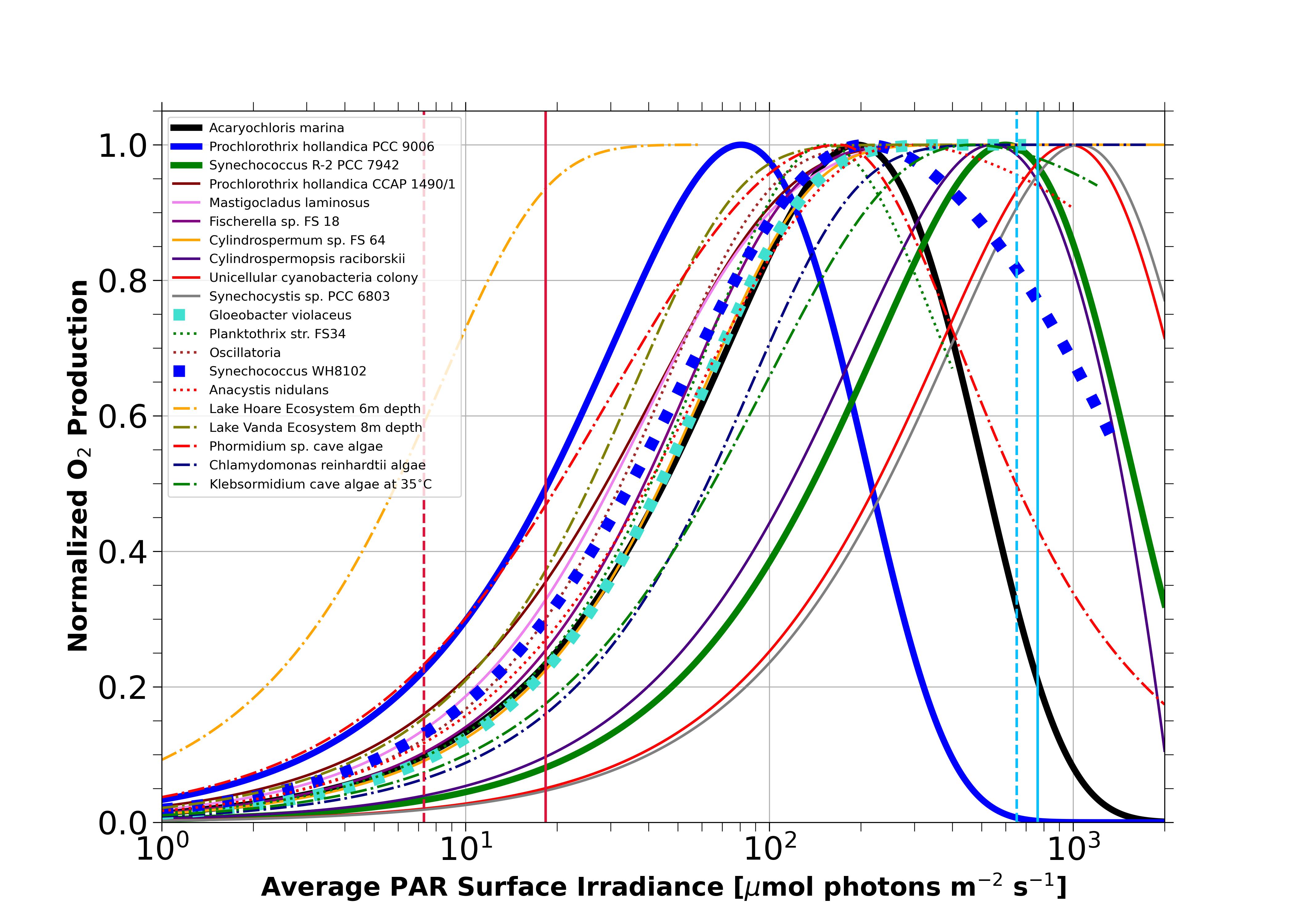}}\\
\subfloat[]{\includegraphics[width = 0.8\textwidth]{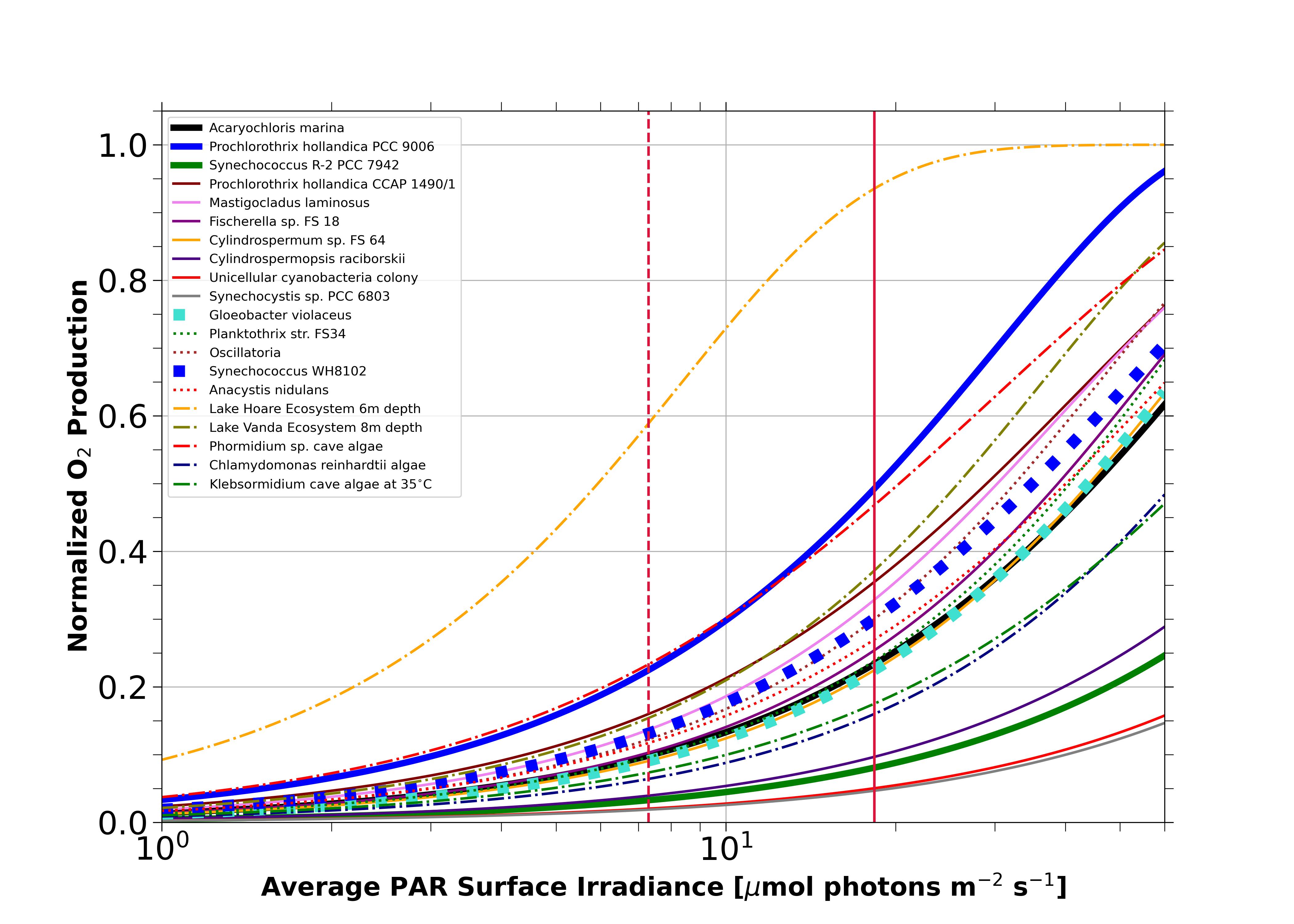}} 
\caption{(a)
Normalized P-I curves showing the relative rate of oxygen production as a function of the light level for various cyanobacteria and related photosynthesizing species. 
Note that the irradiation is shown in logarithmic units.
The more important species for this work are shown in thicker curves.
The mean surface irradiance that the Archean Earth received from the Sun is shown as the vertical cyan lines, for the PAR (dashed)
and the extended PAR (solid).
The vertical red lines shows what a hypothetical Earth would receive 
it if were located in TRAPPIST-1e's orbit.
(b) A zoom-in of the normalized P-I curves for low-light irradiance levels.
}
\label{fig:PI_curve_zoom_out_in}
\end{figure}

\clearpage


\pagebreak
\newpage
\section{Supplemental Information}

In Table~2 we provide the parameters needed to generate the P-I curves shown in Fig.~2.
There are three categories for these curves. The first set include the most relevant species for our work (shown in thicker curves in the Fig.~2). These include the three cyanobacteria
used to estimate the timescales listed in Table~1:
\textit{Acaryochloris marina}, 
\textit{Prochlorothrix hollandica PCC 9006}, and 
\textit{Synechococcus R-2 PCC 7942}
from \cite{ritchie_fitting_2008}.
Other cyanobacteria are
\textit{Prochlorothrix hollandica CCAP 1490/1} 
 (grown at 8 $\mu$mol m\textsuperscript{-2} s\textsuperscript{-1}; 
 \citep{burger-wiersma_functional_1989}), 
\textit{Mastigocladus laminosus} (a filamentous cyanobacteria found in thermal springs;
 \citep{melo_caracterizacion_2021}), 
\textit{Cylindrospermopsis raciborskii} (\citep{OBrien_2009}), 
a microbial mat mix of unicellular cyanobacteria (from the Kaiapoi River of New Zealand 
 (\cite{Dodds_PI_1999}), and 
\textit{Synechocystis sp. strain PCC 6803} (\citep{andersson_fluctuating_2019}). 
For all these cases, the parameters for the P-I curves were explicitly listed in the
associated papers.
Note that for all cases of P-I curves, we omit any dark respiration term because  
the Archean (proto)cyanobacteria we are attempting to match existed in an
anoxic environment (i.e.\ prior to the GOE).

The second set of P-I curves are not cyanobacteria, but are oxygenic photosynthesizers
acclimated to low-light conditions, included for comparison. 
Two are microbial mats containing diatoms, \textit{Leptolyngbya}, and \textit{Oscillatoria} from 
Antarctic lakes of Lake Hoare at 6m depth and from Lake Vanda at 8m depth  (\citep{Hawes_Schwarz_2001}). Also in this set is 
an algae \textit{Chlamydomonas reinhardtii} (grown in CO$_2$-grown cells of 0.1 $\mu$M Fe concentration; \cite{terauchi_trophic_2010}).

The third set of P-I curve parameters are derived from references that did not include the parameters, but only figures. These estimated cases are shown as dashed curves in Fig.~2.
In Table 2, the parameters are preceded by the letter E.
The cyanobacteria in this set include:
\textit{Synechococcus WH8102} (adapted to low-light conditions of 10 $\mu$mol 
  m\textsuperscript{-2} s\textsuperscript{-1}; \citep{Kolodny_2021}), 
\textit{Gloeobacter violaceus} (\citep{KOYAMA2008369}), 
\textit{Planktothrix str.\ FS34} (grown in the absence of H\textsubscript{2}S;
 \citep{Klatt_2015}), and 
\textit{Anacystis nidulans} (grown under 10 $\mu$mol m\textsuperscript{-2} s\textsuperscript{-1}
 light; \citep{samuelsson_susceptibility_1987}). 
Two low-light cave algaes are included: 
\textit{Phormidium sp.} (found in the caves of Frasassi, Italy; \citep{Giordano01092000}),
and 
\textit{Klebsormidium} (based on a photosynthetic rate at 35 degrees C; \citep{futo_ecophysiological_2024}). 
Other P-I curves are from: 
\textit{Fischerella sp.\ FS 18} (grown for 96 hr a pH 9 under DIC limited irradiance of 2 $\mu$mol m\textsuperscript{-2} s\textsuperscript{-1}; \citep{shokravi_growth_2021}) 
and 
\textit{Cylindrospermum sp.\ FS 64} (with salinity 17 mM and pH 9 after 72 h;
\citep{bahavar_acclimation_2022}). For this last case, the parameters were given but
the P-I function was not stated; we assumed the \cite{Jassby_Platt_1976} function.
Likewise we used the \cite{Jassby_Platt_1976} function for
\textit{Oscillatoria} (\citep{martin-clemente_photosynthetic_2022}). 
In the table we have attempted to make the units more homogeneous 
(e.g.\ converting seconds or minutes to hours), 
but for some cases the measured oxygen production rate methodology was significantly different,
and these cannot not readily be scaled to align with the others. However, when plotting P-I curves normalized to their peaks, the oxygen evolution units are scaled out of the problem.

\pagebreak
\newpage
\clearpage

\begin{table}[h]
\caption{Parameters for P-I Curves}
\label{tab:PI_curves_species}
\resizebox{\columnwidth}{!}{%
\begin{tabular}{|>{\itshape}l|l|l|l|}
\hline
\multicolumn{1}{|c|}{\LARGE \textbf{\rule{0pt}{3ex}\rule[-1ex]{0pt}{0pt} Species}} & 
\multicolumn{1}{c|}{\LARGE \textbf{\rule{0pt}{3ex}\rule[-1ex]{0pt}{0pt} Reference}} & 
\multicolumn{1}{c|}{\LARGE \textbf{\rule{0pt}{3ex}\rule[-1ex]{0pt}{0pt} Equation}} & 
\multicolumn{1}{c|}{\LARGE \textbf{\rule{0pt}{3ex}\rule[-1ex]{0pt}{0pt} Parameters}} \\ \hline

Acaryochloris marina & \cite{ritchie_fitting_2008} & Waiting in Line (\cite{ritchie_fitting_2008}) & 
\begin{tabular}[c]{@{}l@{}}$P_{max}$ = 132 $\mu mol \ O_{2} \ (mg \ Chl \ a)^{-1} \ h^{-1}$\\ $k_{w}$ = 0.00515\end{tabular} \\ \hline

Prochlorothrix hollandica PCC 9006 & \cite{ritchie_fitting_2008} & Waiting in Line (\cite{ritchie_fitting_2008}) & 
\begin{tabular}[c]{@{}l@{}}$P_{max}$ = 87.0 $\mu mol \ O_{2} \ (mg \ Chl \ a)^{-1} \ h^{-1}$\\ $k_{w}$ = 0.01240\end{tabular} \\ \hline

Synechococcus R-2 PCC 7942 & \cite{ritchie_fitting_2008} & Waiting in Line (\cite{ritchie_fitting_2008}) & 
\begin{tabular}[c]{@{}l@{}} $P_{max}$ = 313 $\mu mol \ O_{2} \ (mg \ Chl \ a)^{-1} \ h^{-1}$\\ $k_{w}$ = 0.00167\end{tabular} \\ \hline

Prochlorothrix hollandica CCAP 1490/1 & \cite{burger-wiersma_functional_1989} & \cite{dubinsky_light_1986} & 
\begin{tabular}[c]{@{}l@{}}$P_{max}$ = 4.2 $\mu mol \ O_{2} \ (\mu mol \ Chl \ a)^{-1} \ min^{-1}$\\ 
$Chl a_{RC II}$ = 868 $(mol \ Chl \ a)^{-1} \ mol \ RC \ II^{-1}$\\ 
$\sigma_{Chl a}$ = 6.4 $m^{2} \ (mmol \ Chl \ a)^{-1}$\\ 
$\sigma_{RC II}$ = 5555200 $m^{2} \ (mol \ RC \ II)^{-1}$\\ 
$\tau$ = 4.3 $ms$\end{tabular} \\ \hline

Mastigocladus laminosus & \cite{melo_caracterizacion_2021} & \cite{Smith_1936} & 
\begin{tabular}[c]{@{}l@{}}$P_{max}$ = 92 $\mu mol \ O_{2} \ (mg \ Chl \ a)^{-1} \ h^{-1}$\\ 
$\alpha$ = 1.7 $\mu mol \ O_{2} \ (mg \ Chl \ a^{-1}) \ h^{-1}$ $(\mu mol \ photons \ m^{-2} \ s^{-1})^{-1}$\end{tabular} \\ \hline

Fischerella sp. FS 18 & \cite{shokravi_growth_2021} & \cite{Jassby_Platt_1976} & 
\begin{tabular}[c]{@{}l@{}}$P_{max}$ = 268.53 $\mu mol \ O_{2} \ (mg \ Chl \ a)^{-1} \ h^{-1}$\\ 
$\alpha$ = 3.8 $\mu mol \ O_{2} \ (mg \ Chl \ a^{-1}) \ h^{-1}$ $(\mu mol \ photons \ m^{-2} \ s^{-1})^{-1}$\end{tabular} \\ \hline

Cylindrospermum sp. FS 64 & \cite{bahavar_acclimation_2022} & \cite{Jassby_Platt_1976} & 
\begin{tabular}[c]{@{}l@{}}$P_{max}$ = 67.41 $\mu mol \ O_{2} \ (mg \ Chl \ a)^{-1} \ h^{-1}$\\ 
$\alpha$ = 0.84 $\mu mol \ O_{2} \ (mg \ Chl \ a^{-1}) \ h^{-1}$ $(\mu mol \ photons \ m^{-2} \ s^{-1})^{-1}$\end{tabular} \\ \hline

Cylindrospermopsis raciborskii & \cite{OBrien_2009} & \cite{Bright_Walsby_2000} & 
\begin{tabular}[c]{@{}l@{}}$P_{max}$ = 7.5 $\mu g \ C \ (\mu g \ Chl \ a)^{-1} \ h^{-1}$\\ 
$I_{k}$ = 256 $\mu mol \ quanta \ m^{-2} \ s^{-1}$\\ 
$\beta$ = 0.0035 $\mu g \ C \ (\mu g \ Chl \ a)^{-1} \ h^{-1}$ $(\mu mol \ quanta \ m^{-2} \ s^{-1})^{-1}$\end{tabular} \\ \hline

Unicellular cyanobacteria colony & \cite{Dodds_PI_1999} & \cite{Platt1980PhotoinhibitionOP} & 
\begin{tabular}[c]{@{}l@{}}$P_{s}$ = 23.55 $mmol \ O_{2} \ L^{-1} \ min^{-1}$\\ 
$\alpha$ = 0.00023 $mmol \ O_{2} \ L^{-1} \ min^{-1}$ $(\mu mol \ photons \ m^{-2} \ s^{-1})^{-1}$\\ 
$\beta$ = 0.02410 $mmol \ O_{2} \ L^{-1} \ min^{-1}$ $(\mu mol \ photons \ m^{-2} \ s^{-1})^{-1}$\end{tabular} \\ \hline

Synechocystis sp. PCC 6803 & \cite{andersson_fluctuating_2019} & Waiting in Line (\cite{ritchie_fitting_2008}) & 
\begin{tabular}[c]{@{}l@{}}$P_{max}$ = 469 $\mu mol \ O_{2} \ (mg \ Chl \ a)^{-1} \ h^{-1}$\\ 
$k_{w}$ = 1/1050\end{tabular} \\ \hline

Gloeobacter violaceus & \cite{KOYAMA2008369} & \cite{Jassby_Platt_1976} & 
\begin{tabular}[c]{@{}l@{}}$P_{max}$ = 230 $\mu mol \ O_{2} \ (mg \ Chl \ a)^{-1} \ h^{-1}$\\ 
E: $I_{max}$ = 80 $\mu mol \ photons \ m^{-2} \ s^{-1}$\\ 
E: $\alpha$ = 2.875 $\mu mol \ O_{2} \ (mg \ Chl \ a^{-1}) \ h^{-1}$ $(\mu mol \ photons \ m^{-2} \ s^{-1})^{-1}$\end{tabular} \\ \hline

Planktothrix str. FS34 & \cite{Klatt_2015} & \cite{eilers_model_1988} & 
\begin{tabular}[c]{@{}l@{}}E: $P_{max}$ = 0.57 $\mu mol \ O_{2} \ L^{-1} \ s^{-1}$\\ 
E: $I_{opt}$ = 153 $\mu mol \ photons \ m^{-2} \ s^{-1}$\\ 
E: $\alpha$ = 0.00753 $\mu mol \ O_{2} \ L^{-1} \ s^{-1}$ $(\mu mol \ photons \ m^{-2} \ s^{-1})^{-1}$\end{tabular} \\ \hline

Oscillatoria & \cite{martin-clemente_photosynthetic_2022} & \cite{Jassby_Platt_1976} & 
\begin{tabular}[c]{@{}l@{}}$P_{max}$ = 160 $nmol \ O_{2} \ (\mu g \ Chl \ a)^{-1} \ h^{-1}$\\ 
$\alpha$ = 2.7 $nmol \ O_{2} \ (\mu g \ Chl \ a)^{-1} \ h^{-1}$ $(\mu mol \ photons \ m^{-2} \ s^{-1})^{-1}$\end{tabular} \\ \hline

Synechococcus WH8102 & \cite{Kolodny_2021} & \cite{Platt1980PhotoinhibitionOP} & 
\begin{tabular}[c]{@{}l@{}}E: $P_{s}$ = 2100 $\mu mol \ O_{2} \ (mg \ Chl \ a)^{-1} \ h^{-1}$\\ 
E: $\alpha$ = 35 $\mu mol \ O_{2} \ (mg \ Chl \ a^{-1}) \ h^{-1}$ $(\mu mol \ photons \ m^{-2} \ s^{-1})^{-1}$\\ 
E: $\beta$ = 1.11 $\mu mol \ O_{2} \ (mg \ Chl \ a^{-1}) \ h^{-1}$ $(\mu mol \ photons \ m^{-2} \ s^{-1})^{-1}$\end{tabular} \\ \hline

Anacystis nidulans & \cite{samuelsson_susceptibility_1987} & \cite{Platt1980PhotoinhibitionOP} & 
\begin{tabular}[c]{@{}l@{}}E: $P_{s}$ = 330 $\mu mol \ O_{2} \ (mg \ Chl \ a)^{-1} \ h^{-1}$\\ 
E: $P_{max}$ = 400 $\mu mol \ O_{2} \ (mg \ Chl \ a)^{-1} \ h^{-1}$\\ 
E: $I_{k}$ = 75 $\mu mol \ photons \ m^{-2} \ s^{-1}$\\ 
E: $\alpha$ = 5.333 $\mu mol \ O_{2} \ (mg \ Chl \ a^{-1}) \ h^{-1}$ $(\mu mol \ photons \ m^{-2} \ s^{-1})^{-1}$\\ 
E: $\beta$ = 0.05 $\mu mol \ O_{2} \ (mg \ Chl \ a^{-1}) \ h^{-1}$ $(\mu mol \ photons \ m^{-2} \ s^{-1})^{-1}$\end{tabular} \\ \hline

Lake Hoare Ecosystem 6m depth & \cite{Hawes_Schwarz_2001} & \cite{Jassby_Platt_1976} & 
\begin{tabular}[c]{@{}l@{}}$P_{max}$ = 1.34 $\mu g \ cm^{-2} \ h^{-1}$\\ 
$I_{k}$ = 10.8 $\mu mol \ photons \ m^{-2} \ s^{-1}$\\ 
$\alpha$ = 0.12407 $\mu g \ cm^{-2} \ h^{-1}$ $(\mu mol \ photons \ m^{-2} \ s^{-1})^{-1}$\end{tabular} \\ \hline

Lake Vanda Ecosystem 8m depth & \cite{Hawes_Schwarz_2001} & \cite{Jassby_Platt_1976} & 
\begin{tabular}[c]{@{}l@{}}$P_{max}$ = 3.76 $\mu g \ cm^{-2} \ h^{-1}$\\ 
$I_{k}$ = 47 $\mu mol \ photons \ m^{-2} \ s^{-1}$\\ 
$\alpha$ = 0.08 $\mu g \ cm^{-2} \ h^{-1}$ $(\mu mol \ photons \ m^{-2} \ s^{-1})^{-1}$\end{tabular} \\ \hline

Phormidium sp. cave algae & \cite{Giordano01092000} & \cite{eilers_model_1988} & 
\begin{tabular}[c]{@{}l@{}}$P_{max}$ = 38.5 $\mu mol \ O_{2} \ (mg \ Chl \ a)^{-1} \ h^{-1}$\\ 
E: $I_{opt}$ = 170 $\mu mol \ photons \ m^{-2} \ s^{-1}$\\ 
E: $\alpha$ = 1.47 \& 0.47 $\mu mol \ O_{2} \ (mg \ Chl \ a^{-1}) \ h^{-1}$ $(\mu mol \ photons \ m^{-2} \ s^{-1})^{-1}$\end{tabular} \\ \hline

Chlamydomonas reinhardtii algae & \cite{terauchi_trophic_2010} & \cite{Jassby_Platt_1976} & 
\begin{tabular}[c]{@{}l@{}}$P_{max}$ = 3.6 $nmol \ O_{2} \ (nmol \ Chl \ a)^{-1} \ min^{-1}$\\ 
$\alpha$ = 0.013 $nmol \ O_{2} \ (nmol \ Chl \ a)^{-1} \ min^{-1}$ $(\mu mol \ photons \ m^{-2} \ s^{-1})^{-1}$\end{tabular} \\ \hline

Klebsormidium cave algae at 35 degrees C & \cite{futo_ecophysiological_2024} & \cite{Platt1980PhotoinhibitionOP} & 
\begin{tabular}[c]{@{}l@{}}E: $P_{s}$ = 275 $\mu mol \ O_{2} \ (mg \ Chl \ a)^{-1} \ h^{-1}$\\ 
$P_{max}$ = 272 $\mu mol \ O_{2} \ (mg \ Chl \ a)^{-1} \ h^{-1}$\\ 
E: $I_{k}$ = 100 $\mu mol \ photons \ m^{-2} \ s^{-1}$\\ 
E: $\alpha$ = 2.72 $\mu mol \ O_{2} \ (mg \ Chl \ a^{-1}) \ h^{-1}$ $(\mu mol \ photons \ m^{-2} \ s^{-1})^{-1}$\\ 
E: $\beta$ = 0.027 $\mu mol \ O_{2} \ (mg \ Chl \ a^{-1}) \ h^{-1}$ $(\mu mol \ photons \ m^{-2} \ s^{-1})^{-1}$\end{tabular} \\ \hline
\end{tabular}
}
\end{table}

\end{document}